\documentclass[12pt]{article}
\usepackage[T1]{fontenc}
\usepackage[left=1in, top=1in, right=1in, bottom=1in]{geometry}
\usepackage{booktabs} 

\usepackage{amsmath,amssymb,amsthm}
\usepackage{mathtools}
\usepackage{fancyhdr}
\usepackage{graphicx}
\usepackage{enumitem}
\usepackage{aligned-overset}
\usepackage{rotating}
\usepackage{makecell}
\usepackage{booktabs}
\usepackage{multirow}
\usepackage{comment}
\usepackage{derivative}
\usepackage{soul}
\usepackage{setspace}
\usepackage{bbm}
\usepackage{natbib}

\setcitestyle{authoryear}

\usepackage{hyperref}
\hypersetup{colorlinks=true}
\hypersetup{linkcolor=[rgb]{.7,0,0}}
\hypersetup{citecolor=[rgb]{.7,0,0}}
\hypersetup{urlcolor=[rgb]{.7,0,.7}}
\usepackage[nameinlink, capitalise]{cleveref}

\usepackage{autonum}
\usepackage{algorithm, algpseudocodex}
\usepackage{amsmath} 
\usepackage{amssymb}
\usepackage{authblk} 
\usepackage{thmtools, thm-restate}
\declaretheorem[numberwithin=section]{theorem}
\declaretheorem[sibling=theorem, style=definition]{definition}

\declaretheorem[sibling=theorem]{corollary}

\declaretheorem[sibling=theorem]{proposition}

\DeclareMathOperator*{\argmax}{arg\,max}

\DeclareMathOperator*{\E}{\mathbf{E}}

\newcommand{\BR}{\mathrm{BR}}

\newcommand{\Reg}{\mathrm{Reg}}
\newcommand{\ov}{\overline}
\newcommand{\tl}{\widetilde}

\newcommand{\eps}{\varepsilon}

\newcommand{\R}{\mathbb{R}}

\newcommand{\cA}{\mathcal{A}}

\newcommand{\cL}{\mathcal{L}}

\begin{document}
\title{From Best Responses to Learning: Investment Efficiency in Dynamic Environment
\thanks{We are especially grateful to Bart Lipman for his detailed comments on this paper. We also thank Yingkai Li and Weijie Zhong for the valuable discussion. This paper also benefits from numerous audiences at the Stony Brook International Conference on Game Theory. 
}
}
\author{
Ce Li\thanks{Department of Economics, Boston University, email: \texttt{celi@bu.edu}}
\quad
Qianfan Zhang\thanks{Department of Computer Science, Princeton University, email: \texttt{qianfan@princeton.edu}}
\quad
Weiqiang Zheng\thanks{Department of Computer Science, Yale University, email: \texttt{weiqiang.zheng@yale.edu}}
}
\date{}
%
%
%
%

%
\begin{titlepage}
    \clearpage\maketitle
\thispagestyle{empty}
\begin{abstract}


We study the welfare of a mechanism in a dynamic environment where a learning investor can make a costly investment to change her value. In many real-world problems, the common assumption that the investor always makes the best responses, i.e., choosing her utility-maximizing investment option, is unrealistic due to incomplete information in a dynamically evolving environment. 
To address this, we consider an investor who uses a no-regret online learning algorithm to adaptively select investments through repeated interactions with the environment. 
We analyze how the welfare guarantees of approximation allocation algorithms extend from static to dynamic settings when the investor learns rather than best-responds, by studying the approximation ratio for optimal welfare as a measurement of an algorithm’s performance against different benchmarks in the dynamic learning environment. First, we show that the approximation ratio in the static environment remains unchanged in the dynamic environment against the best-in-hindsight benchmark. Second, we provide tight characterizations of the approximation upper and lower bounds relative to a stronger time-varying benchmark. Bridging mechanism design with online learning theory, our work shows how robust welfare guarantees can be maintained even when an agent cannot make best responses but learns their investment strategies in complex, uncertain environments.





\end{abstract}
\end{titlepage}

%
%
\onehalfspacing
\section{Introduction}
\label{sec:intro}
\noindent

Mechanism design is the cornerstone of economic theory,  with applications ranging from resource allocation to online advertising auctions. In many such settings, such as spectrum auctions \cite{doi:10.1073/pnas.1701997114} and online ad auctions \cite{weed2015online, DBLP:conf/nips/BalseiroDMMZ21}, \cite{zuo10.1145/3465456.3467607}, \cite{aggarwal2024auto}, a bidder (investor) could make a costly investment to change her value before participating in the mechanism. An important goal of mechanism design is to incentivize \emph{efficient} investment, that is, the utility-maximizing investment option for the investor is also the investment option that induces optimal social welfare. A fundamental result of \cite{rogerson1992contractual} shows that the Vickrey-Clarke-Groves (VCG) mechanism incentivizes efficient investment. 


However, two obstacles prevent the result in \cite{rogerson1992contractual} from being practical.
First, the VCG mechanism requires computing the optimal allocation, which is computationally intractable in many settings. Although computationally efficient approximation allocation algorithms can achieve near-optimal welfare, when such algorithms are applied in mechanisms that allow investment, the resulting welfare can be devastating~\cite{AKLLM23}. On the positive side, ~\cite{AKLLM23} characterize the sufficient and necessary condition under which the approximation ratio of an allocation algorithm with investment coincides with the approximation ratio of the allocation algorithm without investment.

The second obstacle is that to achieve even approximately optimal social welfare, these results assume the investor chooses her utility-maximizing investment option, \emph{i.e.,} she best responds to the environment. This assumption is often unrealistic in real-world resource allocation problems for various reasons. For example, the investor may have incomplete information to optimize her decision, especially in a dynamically evolving environment. Making the best response in truthful mechanisms requires knowing her threshold price, which depends on other bidders and the mechanism—information the investor may not have. For instance, a bidder in an ad auction typically does not know the values of competing bidders or details about the mechanism. Moreover, the online bidding environment is often rapidly changing, making it difficult for the investor to choose the best option under such variable conditions. A bidder may also have limited cognition, with little computational ability to best respond even with full information ~\cite{ce2023investment}. Consequently, regardless of the computational feasibility of the allocation problem, a bidder's inability to consistently best respond still hinders an allocation algorithm from reaching its (approximately) optimal performance.

While the assumption that the investor makes the best investment decision is unrealistic, a rational investor can instead learn to invest through repeated interactions with the dynamic environment using learning algorithms. The motivation for repeated interactions is that in many real-world applications such as online advertising auctions, the agent participates in a sequence of auctions and sequentially makes investment decisions. Moreover, each interaction with the environment often provides the investor with personal feedback, such as her allocation outcome and her utility. Thus, the investor is able to learn from historical feedback from previous allocation outcomes and utilities to improve her future decision-making. In this paper, we study the social welfare of mechanisms with a \emph{learning} investor in a dynamic environment. In particular, we study the approximation ratio, with regard to the optimal welfare, of an approximation allocation algorithm run on truthful mechanisms with a learning investor.

To relax the unrealistic assumption on best responses, we make a weaker but more realistic assumption: the investor has \emph{no regret}. 
The \emph{regret} is the difference between the investor's actual utility and utility obtained from a certain benchmark option.
By having no regret, the sum of the investor's accumulated utilities over time is asymptotically no less than the utility she would get by choosing the benchmark option. Regret is the measurement of the performance of an online learning algorithm when the actual utilities are brought by the responses chosen by the learning algorithm. Thus, being no-regret describes an ideal performance of an online learning algorithm. In this paper, we additionally make being no-regret a behavioral assumption for an investor who uses such a no-regret online learning algorithm. The assumption of making responses for no regret is supported and motivated by the existence of efficient no-regret learning algorithms (e.g., Exponential Weight \cite{cesa-bianchi_prediction_2006}, the Exp3 algorithm \cite{auer_nonstochastic_2002}) under limited information, as well as no-regret examples in real-world scenarios. For example, \cite{nekipelov2015econometrics} present empirical evidence that bidders' behaviors on Bing are largely consistent with no-regret learning.

Thus, with the more realistic assumption of a learning investor who makes decisions for no regret, a natural question to ask is---\textit{what is the investment efficiency of a truthful mechanism in a dynamic environment with an investor who learns to be no-regret?} Specifically, to what extent will the approximation ratio, with regard to the optimal welfare, of an approximation allocation algorithm be preserved in truthful mechanisms with a learning investor?

In this paper, we allow the investor to make \textit{any} responses by choosing \textit{any} investment decisions. Meanwhile, we open the channel of a dynamic environment for the investor: the investor is able to \textit{learn} to improve her investment decisions through repeated interactions with the exogenous environment. One might expect that the approximation ratio preserved in the dynamic environment where best responses are \textit{not}  required would get worse than the approximation ratio in the one-shot auction where the best response is required. On the contrary, we show that the approximation ratio in the one-shot auction with the best response required can still be preserved in the dynamic environment without the best response needed.

\subsection{An Example}
\noindent

We illustrate our intuition through the example of a knapsack problem \cite{doi:10.1287/opre.5.2.266} and an approximation algorithm \textsc{SmartGreedy} for it. In the knapsack problem, an instance consists of a knapsack size constraint and a set of items, with each item having a value and a corresponding size. The goal is to select items to maximize the total value of the packed items while ensuring the total size of the packed items does not surpass the size of the knapsack. 

The knapsack problem is NP-hard, i.e., it is computationally hard to obtain an exact optimal solution for it. An algorithm to solve the knapsack problem approximately is the \textsc{Greedy} algorithm \cite{doi:10.1287/opre.5.2.266}, which ranks the ratios of values per size of all items in a decreasing order, and keeps packing items from the highest order until the knapsack cannot be packed by an additional item.
The \textsc{SmartGreedy} algorithm \cite{10.5555/1971947} is a standard modification of the \textsc{Greedy} algorithm, which guarantees a $\frac{1}{2}$-approximation to the optimal solution. Specifically, for the \textsc{SmartGreedy} algorithm, we run the \textsc{Greedy} algorithm. Then we compare the result (e.g., welfare) returned by the \textsc{Greedy} algorithm and the most valuable item, and select the one with the higher value.

The knapsack problem with an approximation algorithm (\textsc{Greedy} or \textsc{SmartGreedy}) serves as an example for our intuition about the investment efficiency problem in mechanism design. The rationale is as follows. Each bidder has a value and a size, and his allocation outcome is either being packed or not being packed. The approximation allocation algorithm considered here, either the \textsc{Greedy} algorithm or the \textsc{SmartGreedy} algorithm, is \textit{monotone}: increasing the packed bidder's value does not change his allocation outcome. A monotone allocation algorithm, together with a payment rule, constitutes a truthful mechanism. The payment rule in this example is that the unpacked bidders pay zero, while any packed bidder pays his \textit{threshold price}, i.e., the minimum value that makes him get packed.

Consider the case when the allocation is subject to a knapsack constraint and the seller runs \textsc{SmartGreedy}.
\cite{AKLLM23} show that such an algorithm preserves the worst-case welfare guarantee in the presence of an investor who always makes best responses.
However, the following example (Table \ref{tab:knapsack-example}) illustrates that if the investor selects a slightly suboptimal investment, the resulting welfare can degrade significantly compared to the original welfare given by the best respond.

\begin{table}[h]
    \centering
    \begin{tabular}{l c c c}
        \toprule
        \textbf{Bidder} & $A$ & $B$ & $C$ \\
        \midrule
        \textbf{Value} & $1$ & $1$ & $1$ \\
        \textbf{Size}  & $0.5+\eps$ & $0.5$ & $0.5$ \\
        \bottomrule
    \end{tabular}
    \label{tab:knapsack-example}
    \caption{A knapsack instance with capacity 1. Assume $\eps > 0$.}
    
\end{table}

We observe that in this allocation instance, \textsc{SmartGreedy} packs bidders $B$ and $C$. That leads to a welfare of $2$, which is the optimal outcome. Bidder $A$ is not packed and receives a utility of $0$, since the threshold price of bidder $A$ is $1+2\eps$.

Now suppose that bidder $A$ can invest at a cost of $1$ to increase its value from $1$ to $2+\eps$.
While this investment would ensure $A$ is allocated, it is unprofitable and results in a negative utility of $-\eps$.
Meanwhile, the investment would make \textsc{SmartGreedy} pack $A$ only, which results in a welfare of $1+\eps$.
Compared to the original case where $A$ is not allowed to invest, the welfare with an investor $A$ has decreased by $1-\eps$ even if the investor selects a suboptimal investment that only reduces its utility by $\eps$ only.

Nevertheless, a cautious reader may notice that such a degraded welfare of $1+\eps$ still constitutes an $\frac{1}{2}
$-approximation to the optimal welfare, which may not seem overly concerning. Indeed, such an observation generally holds and we will make use of it later in \cref{sec:result}.

\subsection{Summary of Main Results}
\noindent

For the general allocation problem, each bidder reports his value to the allocation algorithm and receives an allocation outcome determined by the allocation algorithm. One of the bidders, who is an investor, chooses an investment from a finite set of investments, which is a pair consisting of an invested value and a corresponding cost, and reports that invested value to the allocation algorithm.

We consider a dynamic environment with repeated threshold auctions. For each round, the uncertainty of the environment is reflected by the state, which is drawn from a state distribution in that round. The investor has no information about the exogenous environment (e.g., the state, other bidders' values, etc.). She only knows her set of investment options and the total number of rounds before reporting her value to the allocation algorithm. Once the allocation outcomes are assigned to all bidders, the investor receives her payoff in that period. In particular, the investor may rely on an online learning algorithm for making investment decisions over time, which takes the investor's historical payoffs as the input and helps with the investor's future investment decision-making.

Through repeated interactions with the environment, the learning algorithm gets feedback from historical payoffs incurred by past investment decisions and improves its future decision-making based on historical feedback. 
To study the preservation of the approximation ratio of the optimal welfare in the dynamic environment, we measure the performance of an online learning algorithm with the \textit{(expected) regret} incurred by it.
The regret is the difference between the accumulated utility and the utility of the best-in-hindsight investment option.

Our main result is that the approximation ratio of the optimal welfare in a one-shot auction with an investor who is required to always make her best response can still be kept in a dynamic evolving environment with repeated auctions, where the investor's best responses are not required. The general intuition is that the benchmark is the long-run welfare achieved by the best \emph{fixed} investment. The actual welfare suffers a loss from the approximation algorithm, which is implied in the approximation ratio, and also suffers a loss from the actual investment decisions, which constitutes the regret of the learning algorithm.

Additionally, we consider a stronger benchmark, which is the optimal welfare achieved by the best \emph{time-varying} investment options, instead of the best \emph{fixed} investment option. 
We provide a complete characterization of the bounds for approximation ratio in the dynamic environment against the stronger benchmark.


\subsection{Related Literature}
\noindent

There is a line of work on analyzing the investment incentives and efficiency of mechanism design\footnote{Additionally, there is also work on \emph{revenue maximization} in mechanisms when the agents can make investment~\cite{bag1997optimal},\cite{gershkov2021theory} or costly participation, \emph{i.e.,} binary investment~\cite{menezes2000auctions},\cite{celik2009optimal},\cite{gonczarowski2024revenue}.}. We say a mechanism \emph{induces efficient investment} if the \emph{ex-ante} utility-maximizing investment option for an agent is also the \emph{ex-post} socially efficient investment option. A fundamental result of \cite{rogerson1992contractual} shows that the Vickrey-Clarke-Groves (VCG)~\cite{vickrey1961counterspeculation}, \cite{clarke1971multipart}, \cite{groves1973incentives} mechanism induces efficient investment. \cite{bergemann2002information} extend this result to a mechanism design setting with uncertainty, where each agent can invest in information before participating in the mechanism. For investment incentives in auction settings, \cite{king1992investment} and \cite{arozamena2004investment} demonstrate that second-price auctions induce efficient investment while first-price auctions result in inefficient underinvestment. \cite{hatfield2018strategy} extends these results by showing that a mechanism induces efficient investment if and only if it is \emph{ex-post} efficient and strategy-proof. The work of \cite{hatfield2018strategy} also characterizes how the additive error bounds of a mechanism fail to be efficient or strategy-proof relates to its absolute loss in inefficient investment. 

The most related work to our work is \cite{AKLLM23}, which studies the investment efficiency of a mechanism that uses an \emph{approximately} efficient allocation algorithm. \cite{AKLLM23} characterize the sufficient and necessary conditions on the algorithm under which the approximation ratios for the allocation and investment are equal. All the work above only considers the case when the investor \emph{best responds}\footnote{When multiple agents can invest, these works assume they best respond to each other, and their strategies form a (Bayesian) Nash equilibrium.} to the environment, \emph{i.e.,} chooses the utility-maximization investment option. However, to choose the best investment option \emph{ex-ante}, the investor needs substantial information about the mechanism and other agents. Unlike these results, we focus on a setting where the investor may not have information about the mechanism or the environment. Instead, the unknown environment is dynamically evolving, and the investor must learn to invest by repeated interaction. 


Our work also contributes to the literature on games with learning agents. This line of work is motivated by realistic settings where common assumptions fail to hold. In particular, \cite{nekipelov2015econometrics} relaxed the Nash equilibrium assumption and studied how to perform inference of bidder values under the weaker assumption that bidders are using no-regret learning. \cite{camara2020mechanisms}, \cite{collina2024efficient} model agent's long-run behavior with the no-regret behavior assumption that relaxes the common prior assumption. Starting from \cite{braverman2018selling}, there is a line of works on strategizing against a no-regret/learning agent in various settings including Stackelberg games~\cite{deng2019strategizing}, Bayesian games~\cite{mansour2022strategizing}, auctions~\cite{cai2023selling}, \cite{rubinstein2024strategizing}, contract design~\cite{guruganesh2024contracting}, information design~\cite{jain2024calibrated},\cite{yang2024computational},\cite{lin2025informationdesignunknownprior}, and general principle-agent problems~\cite{lin2025generalized}.


\section{Model}
\label{sec:model}


\subsection{Allocation and Approximation Algorithm}
\noindent

We consider a finite set of bidders $N$ and a finite set of outcomes $O$. For example, the set of outcomes $O$ can be $\{0,1\}$ to denote binary allocation outcomes $\{\text{getting packed}, \text{not getting packed}\}$ for bidders in the knapsack problem. An \textbf{allocation instance} $(v,A)$ consists of a value profile $v=(v_{n,o})_{n \in N, o \in O}$, where $v_{n,o}\in \R_{\ge 0}$ denotes the value of bidder $n$ whose allocation outcome is $o$, and a set of feasible allocations $A \subseteq O^N$. The \textbf{optimal welfare} at an allocation instance $(v,A)$ is
$W^*(v,A)=\max_{a \in A} \sum_{n} v_{n} \cdot a_n,$
where we abuse notation and treat $a_n$ as an indicator vector over outcomes $O$.

An \textbf{allocation problem} $\Omega$ is a collection of allocation instances. We assume value profiles in $\Omega$ have a product structure, i.e., $\{v: (v,A) \in \Omega\}=\prod_{n \in N} \prod_{o \in O} V_{n,o}$. 
In the following, we will fix the allocation problem $\Omega$ we consider and further denote the value $V_n=\prod_{o \in O} V_{n,o}$ for each bidder $n \in N$ and the value $V=\prod_{n \in N} V_n$ for all bidders.

An \textbf{allocation algorithm} $x$ for an allocation problem $\Omega$ is a function that maps every instance $(v,A) \in\Omega$ to a feasible allocation $x(v,A) \in A$. We further denote the outcome for bidder $n$ under $x$ by $x_n(v,A)$. The \textbf{welfare} of algorithm $x$ at instance $(v,A)$ is
        $W_x(v,A)=\sum_{n} v_{n} \cdot x_n(v,A),$
where we abuse notation and treat $x_n(v,A)$ as an indicator vector over outcomes $O$. For some $\beta \in [0,1]$, an allocation algorithm $x$ is a \textbf{$\beta$-approximation for allocation}, if for every instance $(v,A) \in \Omega$, 
        $W_x(v,A) \ge \beta W^*(v,A).$


A \textbf{mechanism} $(x,p)$ consists of an allocation algorithm $x$ and a payment rule $p$ that maps every reported instance $(\hat{v},A)$ into a feasible allocation $x(\hat{v},A) \in A$ and a payment profile $p(\hat{v},A) \in \R^N$. A mechanism $(x,p)$ is \textbf{truthful} for an allocation problem $\Omega$ if for all instances $(v,A) \in \Omega$ and all $\hat v_n \in V_n$,
        $v_n \cdot x_n(v,A)-p_n(v,A) \ge v_n \cdot x_n(\hat v_n, v_{-n},A) - p_n(\hat v_n, v_{-n},A).$
An allocation algorithm $x$ is \textbf{weakly monotone} if $[v_n' - v_n] \cdot [x_n(v_n', v_{-n},A)-x_n(v,A)] \ge 0.$
An allocation algorithm $x$ is weakly monotone if and only if $(x,p)$ is truthful.
\begin{theorem}[\cite{lavi2003towards,saks2005weak}]
    When value profiles in $\Omega$ have a product structure, an algorithm $x$ is weakly monotone if and only if $(x,p)$ is truthful for some payment rule $p$.
\end{theorem}

\subsection{Static Investment Environment}
\noindent

To construct the static (one-shot) investment environment, we model the \textit{ex ante} uncertainty the investor faces by a finite set of states $S$. For example, such uncertainty may be that the investor does not know the mechanism that is running. The uncertainty may also be that the investor does not know the values of the other bidders. By being static, we mean that the auction is run only once. In this way, there will be no repeated interactions from the investor with the environment, so learning in such a static environment is impossible. 

Formally\footnote{Many of our notations are borrowed from \cite{AKLLM23}.}, for the static environment, we fix a finite set of states $S$ and an investor $\iota \in N$. We also fix a truthful mechanism $(x,p)$ for the allocation problem $\Omega$.
We define a \textbf{(static) investment instance} $\ov{\omega}=(g,I,\nu_{-\iota},\cA)$ to be a tuple that consists of:
\begin{itemize}
    \item a state distribution $g\in \Delta S$,
    \item a finite set of investments $I$, where an \emph{investment}  $(\nu_\iota,c)$ consists of a function from states to bidder $\iota$'s value $\nu_\iota: S \to V_\iota$ and a cost $c\in \R$ 
    (we require that $I$ contains at least one pair $(\nu_\iota,c)$ with $c=0$),
    \item a function from states to other bidders' values, $\nu_{-\iota}:S \to V_{-\iota}$,
    \item a correspondence from states to feasible allocations, $\cA: S \rightrightarrows O^N$. 
\end{itemize}

We assume that the resulting allocation instance $(\nu(s),\cA(s)) \in \Omega$ belongs to the allocation problem for each state $s \in S$ and that the investor's investment $(\nu_\iota,c)\in I$ is taken from her finite set of investments $I$. When running the truthful mechanism $(x,p)$ at instance $\ov{\omega}$, we define the following performance quantities.

The \textbf{utility} of the investor\footnote{Note that the investor's utility also depends on the mechanism $(x,p)$ but we omit them for the ease of notation. } $\iota$ who chooses an investment $(\nu_\iota,c)$ at instance $\ov{\omega}$ given state $s\in S$ is computed by
        $u(\ov{\omega}, s, \nu_\iota,c) = \nu_\iota(s) \cdot x_\iota(\nu(s),\cA(s)) - p_\iota(\nu(s),\cA(s))-c.$

The \textbf{welfare} achieved by an allocation algorithm $x$ at instance $\ov{\omega}$ given state $s \in S$ when investor $\iota$ chooses $(\nu_\iota,c)\in I$ is 
        $\ov{W}_x(\ov{\omega},s,\nu_\iota,c)= W_x(\nu(s),\cA(s)) - c.$
The \textbf{optimal welfare} at instance $\ov{\omega}$ given state $s \in S$ when investor $\iota$ chooses $(\nu_\iota,c)\in I$ is
        $\ov{W}^*(\ov{\omega}, s, \nu_\iota,c)= W^*(\nu(s),\cA(s)) - c.$
        
With a slight abuse of notation, we further define the following expectations based on the randomness over states. Specifically, the \textbf{expected utility} of the investor is by denoted by $u(\ov{\omega}, \nu_\iota,c) = \E_{s\sim g}[ u(\ov{\omega}, s,\nu_\iota,c)].$ The \textbf{expected welfare} obtained by an approximation allocation algorithm $x$ is $\ov{W}_x(\ov{\omega},\nu_\iota,c)=\E_{s\sim g}[\ov{W}_x(\ov{\omega},s,\nu_\iota,c)].$ We also define the \textbf{expected optimal welfare} by $\ov{W}^*(\ov{\omega},\nu_\iota,c)=\E_{s\sim g}[\ov{W}^*(\ov{\omega},s,\nu_\iota,c)].$

The \textbf{best-response investment} for investor $\iota$ at instance $\ov{\omega}$ is $\BR(\ov{\omega})=\argmax_{(\nu_\iota,c)\in I} u(\ov{\omega},\nu_\iota,c).$
We define an algorithm $x$ to be a \textbf{$\beta$-approximation for (static) investment}, if, for every instance $\ov{\omega}$, 
        $$\min_{(\nu_\iota,c)\in \BR(\ov{\omega})} \ov{W}_x(\ov{\omega},\nu_\iota,c) \ge \beta \max_{(\nu_\iota,c)\in I} \ov{W}^*(\ov{\omega},\nu_\iota,c).$$
Note that in the static setting, we assume the investor $\iota$ always chooses a best-response investment, although it might be the one that gives the worst welfare among them.

Since the investment is made with a certain cost by the investor, the approximation ratio for allocation will be weakly decreased when the same allocation algorithm is run.
\cite{AKLLM23} provide a tractable way to assess whether an approximation ratio for allocation remains the same when the investment is allowed to be made, in comparison with the situation where no investment is allowed.

\begin{definition}[\cite{AKLLM23}]
An algorithm $x$ \textbf{excludes confirming negative externalities (is ``XCONE'')} if for any instance $(v,A)$ and any change from $v_n$ to $v_n'$ that confirms $x_n(v,A)$, 
        $$\sum_m [v_m \cdot [x_m(v_n', v_{-n},A)-x_m(v,A)]] \ge 0,$$
where a change from $v_n$ to $v_n'$ \emph{confirms} an outcome $o'$ if $[v_n'-v_n]\cdot [o'-o] \ge 0$ for all outcomes $o$.
\end{definition}

\begin{theorem}[\cite{AKLLM23}]\label{thm:allocation-to-investment-approx}
For any weakly monotone algorithm $x$ and any $\beta \in [0,1]$, if $x$ is XCONE and a $\beta$-approximation for allocation, then $x$ is also $\beta$-approximation for (static) investment.
\end{theorem}


The XCONE property is satisfied by several commonly used approximation algorithms. For example, both the \textsc{Greedy} algorithm and the \textsc{SmartGreedy} algorithm for the knapsack problem are XCONE. Moreover, beyond constant approximation, \cite{AKLLM23} propose a modified ``fully polynomial time approximation scheme'' (FPTAS) of \cite{BKV05} that is also XCONE. The existence of XCONE allocation algorithms in the static investment environment presents natural candidates for consideration in dynamic environments, thereby motivating us to study whether the performance guarantees of those XCONE allocation algorithms in the static setting can be preserved in dynamic environments.

\subsection{Dynamic Learning Environment}

\noindent

In the dynamic learning environment, we keep fixed the set of bidders $N$, the outcomes $O$, the allocation problem $\Omega$, a truthful mechanism $(x,p)$ for $\Omega$, the set of states $S$, and an investor $\iota \in N$.

We define a \textbf{dynamic investment instance} to be a tuple $\tl{\omega}^T=(T,I, \{ (g^{(t)}, \nu_{-\iota}^{(t)},\cA^{(t)})\}_{t\in[T]})$ that consists of:
\begin{enumerate}
    \item a fixed set of investments $I$,
    \item the number of rounds $T$,
    \item a state distribution $g^{(t)} \in \Delta S$ for each round $t \in [T]$,
    \item a function from states to the values of other bidders, $\nu_{-\iota}^{(t)}: S \to V_{-\iota}$ for each round $t \in [T]$,
    \item a correspondence from states to feasible allocations, $\cA^{(t)}:  S \rightrightarrows O^N$ for each round $t \in [T]$.
\end{enumerate}

Given a defined dynamic investment instance $\tl{\omega}^T$, the investor $\iota$ \textit{knows} his fixed set of investments $I$ and the length of the learning process $T$. He  has \textit{no} information about the state distribution $g^{(t)} \in \Delta S$, other bidders' values $\nu_{-\iota}^{(t)}$, and feasible allocations $\cA^{(t)}$ in each round $t$.

The dynamic investment instance represents a single-agent game for the investor $\iota$. Sequentially, for each round $t=1,2,\ldots,T$, the investor $\iota$ participates in the truthful mechanism $(x,p)$ at an investment instance $\ov{\omega}^{(t)} = (g^{(t)}, I, \nu_{-\iota}^{(t)}, \cA^{(t)})$ in the following order.
    \begin{enumerate}
            \item A (random variable) state $s^{(t)}$ is drawn from a state distribution $g^{(t)}$.
            \item The investor picks an investment $(\nu_\iota^{(t)},c^{(t)}) \in I$ without knowing the state distribution $g^{(t)}$, the mapping from states to the other bidders' values $\nu_{-\iota}^{(t)}$, or the correspondence from states to feasible allocations $\cA^{(t)}$ for the round $t$.
            \item The investor receives his utility $u^{(t)}=u(\ov{\omega}^{(t)},s^{(t)},\nu_\iota^{(t)},c^{(t)})$.
    \end{enumerate}
    
We remark that the utility $u^{(t)}$ is the only feedback the investor has access to at round $t$, which is known as the \emph{bandit feedback} in the online learning literature (e.g., \cite{slivkins2024introductionmultiarmedbandits}, \cite{lattimore2020bandit}). In other words, the decision of the investor at round $t$ can only depend on the investor's utilities $u^{(1)}, u^{(2)},\ldots, u^{(t-1)}$ in the past rounds, the fixed set of investments $I$, and its internal randomness.

Instead of picking an investment by himself, the investor may choose an investment by using an \textbf{online learning algorithm} $\cL$, which takes the set of investments $I$ and utilities $u^{(1)},u^{(2)},\ldots,u^{(t-1)}$ as input and outputs an investment $(\nu_\iota^{(t)},c^{(t)}) \in I$ at each round $t \in [T]$.


To measure the performance of an online learning algorithm $\cL$, we define the \textbf{(expected) regret} of algorithm $\cL$ to be the difference between the actual utilities of the investor incurred by $\cL$ and the best-in-hindsight utility.
\begin{definition}
   The \textbf{(expected) regret} of an online learning algorithm $\cL$ at instance $\tl{\omega}^T$ is $$\Reg(\tl{\omega}^T,\cL)=\E_{s^{(t)}\sim g^{(t)},\cL}\left[\max_{(\nu_\iota,c)\in I} \sum_{t=1}^T u(\ov{\omega}^{(t)},s^{(t)},\nu_\iota,c) - \sum_{t=1}^T u(\ov{\omega}^{(t)},s^{(t)},\nu_\iota^{(t)},c^{(t)})\right],$$
    where the expectation is taken with respect to all $s^{(t)}$ and the internal randomness of $\cL$, and $\nu_\iota^{(t)},c^{(t)}$ is determined by $\cL$ at round $t$.
\end{definition}

An online learning algorithm is \textit{no-regret} if the regret incurred by it grows sublinearly in time (or, the time-averaging regret converges to zero as time goes to infinity). Specifically, in the dynamic learning environment for investment, we call an online learning algorithm $\cL$ \textbf{no-regret} if, for every dynamic investment instance $\tl{\omega}^T=(T,I,\cdot)$, 
    $\Reg(\tl{\omega}^T,\cL) \le f_{\Omega,I}(T)$
for some fixed function $f_{\Omega,I}(T) = o(T)$. Without loss of generality, we assume the investor's utility is in the range of [0,1]. Many online learning algorithms developed for the multi-armed bandit problem are no-regret for dynamic investment~\cite{bubeck2012regret,lattimore2020bandit}. We present the classic EXP3 algorithm~\cite{auer_nonstochastic_2002} tailored for our setting below for completeness.
    \begin{algorithm}
    \caption{EXP3 for dynamic investment}\label{alg:EXP3}
        \begin{algorithmic}
            \State \textbf{Input:} $T\ge 1, I$ 
            \State \textbf{Initialization:} $q^{(1)}[(\nu_\iota, c)] = 1$ for all $ (\nu_\iota,c) \in I$, $\gamma = \min\{1, \sqrt{\frac{|I|\log |I|}{(e-1)T}}\}$
            \For{$t=1, 2, \ldots, T$}
            \State Set $p^{(t)}[(\nu_\iota, c)] = (1-\gamma)\frac{q^{(t)}[(\nu_\iota, c)]}{\sum_{(\nu_\iota', c')\in I}q^{(t)}[(\nu_\iota', c')]} + \frac{\gamma}{|I|}$ for all $i \in I$.
            \State Draw and choose investment $(\nu_\iota^{(t)}, c^{(t)})\in I$ according to probability $p^{(t)}$.
            \State Receive utility $u^{(t) } :=u(\ov{\omega}^{(t)},s^{(t)},\nu_\iota^{(t)},c^{(t)})$
            \State Set $q^{(t+1)}[(\nu_\iota^{(t)}, c^{(t)})] = q^{(t)}[(\nu_\iota^{(t)}, c^{(t)})] \cdot \exp\left( \frac{\gamma u^{(t)}}{|I| p^{(t)}[(\nu_\iota^{(t)}, c^{(t)})]} \right)$ and $q^{(t+1)}[(\nu_\iota, c)] = q^{(t)}[(\nu_\iota, c)]$ for all $(\nu_\iota, c) \ne (\nu_\iota^{(t)}, c^{(t)})$.
            \EndFor
        \end{algorithmic}
    \end{algorithm}

    
\begin{theorem}[\cite{auer_nonstochastic_2002}]
\label{thm:no-regret-learning}
        \Cref{alg:EXP3} guarantees $\Reg(\tl{\omega}^T, \cL)  = O(\sqrt{T|I|\log|I|})$.
\end{theorem}

Then, we characterize the measure for the performance of an allocation algorithm $x$ in such a dynamic learning environment. The \textbf{(expected) welfare} of an allocation algorithm $x$ at instance $\tl{\omega}^T$ when the investor is using an online learning algorithm $\cL$ is defined as
        $$\tl{W}_x(\tl{\omega}^T,\cL)=\E_{s^{(t)}\sim g^{(t)},\cL}\left[\sum_{t=1}^T \ov{W}_x(\ov{\omega}^{(t)}, s^{(t)},\nu_\iota^{(t)},c^{(t)})\right].$$
The \textbf{optimal (expected) welfare} at a dynamic investment instance $\tl{\omega}^T$ is
        $$\tl{W}^*(\tl{\omega}^T)=\max_{(\nu_\iota,c)\in I} \sum_{t=1}^T \ov{W}^*(\ov{\omega}^{(t)},\nu_\iota,c).$$
An allocation algorithm $x$ is a \textbf{$\beta$-approximation for dynamic investment}, if for every dynamic investment instance $\tl{\omega}^T$ and an online learning algorithm $\cL$,
\begin{align}
\label{def:beta-dynamic}
 \tl{W}_x(\tl{\omega}^T,\cL) \ge \beta \tl{W}^*(\tl{\omega}^T) - \Reg(\tl{\omega}^T,\cL).   
\end{align}
Specifically, when the online learning algorithm $\cL$ is \textit{no-regret}, \cref{def:beta-dynamic} further implies
        $\tl{W}_x(\tl{\omega}^T,\cL) \ge \beta \tl{W}^*(\tl{\omega}^T) - o(T),$
where $o(T)$ denotes that $\lim_{T \rightarrow \infty}\frac{\Reg(\tl{\omega}^T,\cL)}{T}=0$.


\section{Main Results}
\label{sec:result}
\noindent

We characterize the approximation ratios in the dynamic learning environment for different benchmarks. In Section \ref{sec:approximation-dynamic-investment}, we study approximation for dynamic investment. We show that the approximation ratio of an allocation algorithm for static investment is preserved for dynamic investment.
In Section \ref{sec:stronger benchmark}, we consider a stronger time-varying benchmark and introduce the notion of \textit{approximation for strongly dynamic investment} for such a stronger benchmark. We show that an algorithm's approximation ratio is affected by the number of investment options. We then characterize the lower and upper bounds for the approximation ratio and show that they match. 

\subsection{Approximation for Dynamic Investment}
\label{sec:approximation-dynamic-investment}

\noindent

This section establishes that the approximation guarantees of weakly monotone algorithms in the static investment setting extend to the dynamic setting, despite the additional complexity introduced by a learning investor. Specifically, we characterize the extension in the following theorem.

\begin{theorem}\label{thm:beta-approx-dynamic}
    For any weakly monotone algorithm $x$ and any $\beta \in [0,1]$, if $x$ is $\beta$-approximation for static investment, then $x$ is also $\beta$-approximation for dynamic investment.
\end{theorem}

Combining \cref{thm:beta-approx-dynamic} with \cref{thm:allocation-to-investment-approx}, we can immediately obtain \cref{cor:XCONE-to-dynamic-efficiency}, which can be applied to a broad class of approximation allocation algorithms, such as the \textsc{Greedy} algorithm and the \textsc{SmartGreedy} algorithm for the knapsack problem. 
What's more, for the knapsack problem, there exists a fully polynomial time approximation scheme (FPTAS) that is XCONE and gives a nearly optimal approximation ratio of $1-\varepsilon$ for any $\varepsilon>0$ \cite{BKV05,AKLLM23}.

\begin{corollary}\label{cor:XCONE-to-dynamic-efficiency}
    For any weakly monotone algorithm $x$ and any $\beta \in [0,1]$, if $x$ is XCONE and a $\beta$-approximation for allocation, then $x$ is also $\beta$-approximation for dynamic investment.
\end{corollary}

The key challenge in the dynamic investment model is that welfare loss arises not only from the approximation algorithm $x$ but also from the investor: the investor employs an online learning algorithm for her decision-making, but the investments chosen by the online learning algorithm may be suboptimal, even when the online learning algorithm is no-regret.
Such a gap between the investor's actual utility and her optimal utility in hindsight is characterized by the investor's regret, and a small regret of $o(T)$ is achievable in general (\cref{thm:no-regret-learning}).

It turns out that, if the approximation algorithm $x$ is $\beta$-approximation for static investment, whenever the investor selects an investment $(\nu_\iota,c)$ with a utility gap of $\delta$ compared to the best response, the resulting welfare remains at least $\beta$-fraction of the optimal welfare minus $\delta$.
To see this, consider a new instance where $(\nu_\iota,c)$ is replaced by $(\nu_\iota,c-\delta)$, effectively lowering the cost while preserving the value.
Since the allocation algorithm $x$ cannot distinguish $(\nu_\iota,c-\delta)$ from $(\nu_\iota,c)$, the allocation remains unchanged if the investor selects $(\nu_\iota,c-\delta)$ in the new instance.
Consequently, the modified investment becomes a best response and achieves a $\beta$-approximation of the optimal welfare by \cref{thm:allocation-to-investment-approx}.
This, in turn, establishes our original claim, as $(\nu_\iota,c-\delta)$ yields a welfare that is exactly $\delta$ larger than $(\nu_\iota,c)$.

Using this observation, we prove \cref{thm:beta-approx-dynamic} by comparing the investor's choices $(\nu_\iota^{(t)},c^{(t)})$ with the optimal investment $(\nu_\iota^*,c^*)$ and relating their welfare differences to utility differences.

\begin{proof}[Proof of \cref{thm:beta-approx-dynamic}]
    Let $\tl{\omega}^T=(T,I, \{ (\sigma^{(t)}, \nu_{-\iota}^{(t)},\cA^{(t)})\}_{t\in[T]})$ be any dynamic investment instance.
    Let $(\nu_\iota^*,c^*) \in I$ be the investment that achieves the optimal welfare $\tl{W}^*(\tl{\omega}^T)$, i.e.,
    $$(\nu_\iota^*,c^*) = \argmax_{(\nu_\iota,c)\in I} \sum_{t=1}^T \ov{W}^*(\ov{\omega}^{(t)},\nu_\iota,c).$$

    For each round $t \in [T]$ and the static investment instance $\ov{\omega}^{(t)}=(g^{(t)},I,\nu_{-\iota}^{(t)},\cA^{(t)})$, let the random variable $\delta^{(t)}$ be the utility difference between $(\nu_\iota^*,c^*)$ and $(\nu_\iota^{(t)},c^{(t)})$, i.e., 
    $\delta^{(t)} = u(\ov{\omega}^{(t)}, s^{(t)}, \nu_\iota^*, c^*) - u(\ov{\omega}^{(t)}, s^{(t)}, \nu_\iota^{(t)}, c^{(t)}).$
    Also note that, by definition of regret,
    \begin{equation}
    \E_{s^{(t)}\sim g^{(t)},\cL}\left[ \sum_{t=1}^T \delta^{(t)} \right] \le \Reg(\tl{\omega}^T,\cL). \label{eq:sum-of-delta-leq-regret}
    \end{equation}
    
    Consider a modified instance $\ov{\omega}^{(t)}_\circ=(g^{(t)}_\circ,I^{(t)}_\circ,\nu_{-\iota}^{(t)},\cA^{(t)})$ (which depends on $s^{(t)}$,  $(\nu_\iota^{(t)},c^{(t)})$, $\delta^{(t)}$) where
    \begin{itemize}
        \item $g^{(t)}_\circ$ is a degenerate distribution that gives state $s^{(t)}$ with probability $1$.
        \item $I^{(t)}_\circ= \{(\nu_\iota^*,c^*), (\nu_\iota^{(t)},c^{(t)}-\delta^{(t)})\}$,
    i.e., keeping $(\nu_\iota^*,c^*)$ unmodified, replacing $(\nu_\iota^{(t)},c^{(t)})$ with another investment $(\nu_\iota^{(t)},c^{(t)}-\delta^{(t)} )$ with a different cost\footnote{It is possible that $c^{(t)}-\delta^{(t)} < 0$, making $(\nu_\iota^{(t)},c^{(t)}-\delta^{(t)})$ a disinvestment, which is also considered valid.}, and discarding all other investments.
    \end{itemize}
    In the new instance $\ov{\omega}^{(t)}_\circ$, observe that for investment $(\nu_\iota^*,c^*)$, the corresponding utility, welfare, and optimal welfare all remain unchanged (we omit state $s^{(t)}$ for $\ov{\omega}^{(t)}_\circ$ as $g^{(t)}_\circ$ being degenerate):
    \begin{align}
        u(\ov{\omega}^{(t)}_\circ,\nu_\iota^*,c^*)&=u(\ov{\omega}^{(t)},s^{(t)},\nu_\iota^*,c^*), \label{eq:utility-unchanged}\\
        \ov{W}_x(\ov{\omega}^{(t)}_\circ,\nu_\iota^*,c^*)&=\ov{W}_x(\ov{\omega}^{(t)},s^{(t)},\nu_\iota^*,c^*), \label{eq:welfare-unchanged}\\
        \ov{W}^*(\ov{\omega}^{(t)}_\circ,\nu_\iota^*,c^*)&=\ov{W}^*(\ov{\omega}^{(t)},s^{(t)},\nu_\iota^*,c^*). \label{eq:opt-welfare-unchanged}\
    \end{align}
    Meanwhile, for investment $(\nu_\iota^{(t)},c^{(t)}-\delta^{(t)})$, its corresponding utility, welfare, and optimal welfare at state $s^{(t)}$ have all increased by $\delta^{(t)}$:
    \begin{align}
        u(\ov{\omega}^{(t)}_\circ,\nu_\iota^{(t)},c^{(t)}-\delta^{(t)})&=u(\ov{\omega}^{(t)},s^{(t)},\nu_\iota^{(t)},c^{(t)})+\delta^{(t)}, \label{eq:utility-by-delta}\\
        \ov{W}_x(\ov{\omega}^{(t)}_\circ,\nu_\iota^{(t)},c^{(t)}-\delta^{(t)})&=\ov{W}_x(\ov{\omega}^{(t)},s^{(t)},\nu_\iota^{(t)},c^{(t)})+\delta^{(t)}, \label{eq:welfare-by-delta}\\
        \ov{W}^*(\ov{\omega}^{(t)}_\circ,\nu_\iota^{(t)},c^{(t)}-\delta^{(t)})&=\ov{W}^*(\ov{\omega}^{(t)},s^{(t)},\nu_\iota^{(t)},c^{(t)})+\delta^{(t)}. \label{eq:opt-welfare-by-delta}
    \end{align}
    Combining \cref{eq:utility-unchanged} and \cref{eq:utility-by-delta} with the definition of $\delta^{(t)}$, we can obtain
    $$u(\ov{\omega}^{(t)}_\circ,\nu_\iota^{(t)},c^{(t)}-\delta^{(t)}) = u(\ov{\omega}^{(t)}_\circ,\nu_\iota^*,c^*),$$
    which further implies both $(\nu_\iota^{(t)},c^{(t)}-\delta^{(t)}) \in \BR(\ov{\omega}^{(t)}_\circ)$ and $(\nu_\iota^*,c^*) \in \BR(\ov{\omega}^{(t)}_\circ)$.
    Since $(x,p)$ is a $\beta$-approximation for static investment, running the mechanism $(x,p)$ at instance $\ov{\omega}^{(t)}_\circ$ when the investor $\iota$ chooses $(\nu_\iota^{(t)},c^{(t)}-\delta^{(t)})$ yields welfare
    $$\ov{W}_x(\ov{\omega}^{(t)}_\circ,\nu_\iota^{(t)},c^{(t)}-\delta^{(t)}) \ge \beta \max_{(\nu_\iota,c) \in I^{(t)}_\circ} \ov{W}^*(\ov{\omega}^{(t)}_\circ,\nu_\iota,c) \ge \beta \ov{W}^*(\ov{\omega}^{(t)}_\circ,\nu_\iota^*,c^*).$$
    By \cref{eq:opt-welfare-unchanged} and \cref{eq:welfare-by-delta}, we can conclude that
    \begin{equation}
      \ov{W}_x(\ov{\omega}^{(t)},s^{(t)},\nu_\iota^{(t)},c^{(t)}) \ge \beta \ov{W}^*(\ov{\omega}^{(t)},s^{(t)},\nu_\iota^*,c^*) - \delta^{(t)}.  \label{eq:beta-approx-by-delta}
    \end{equation}
    And the welfare of algorithm $x$ at instance $\tl{\omega}^T$ when the investor is using $\cL$ is
    \begin{align}
    \tl{W}_x(\tl{\omega}^T,\cL)
    &= \E_{s^{(t)}\sim g^{(t)},\cL}\left[\sum_{t=1}^T \ov{W}_x(\ov{\omega}^{(t)}, s^{(t)},\nu_\iota^{(t)},c^{(t)})\right] \\
    &\ge \E_{s^{(t)}\sim g^{(t)},\cL}\left[\sum_{t=1}^T \beta \ov{W}^*(\ov{\omega}^{(t)},s^{(t)},\nu_\iota^*,c^*) - \delta^{(t)} \right] \tag{\cref{eq:beta-approx-by-delta}}\\
    &= \beta \sum_{t=1}^T  \ov{W}^*(\ov{\omega}^{(t)},\nu_\iota^*,c^*) - \E_{s^{(t)}\sim g^{(t)},\cL}\left[ \sum_{t=1}^T \delta^{(t)} \right]\\
    &\ge \beta \tl{W}^*(\tl{\omega}^T) - \Reg(\tl{\omega}^T,\cL), \tag{\cref{eq:sum-of-delta-leq-regret}}
    \end{align}
    which concludes our proof.
\end{proof}



\subsection{On Approximation of a Stronger Benchmark} 
\label{sec:stronger benchmark}
\noindent

In evaluating the performance of an approximation allocation algorithm when an investor uses an online learning algorithm for her decision-making, one might be interested in a stronger benchmark, a time-varying benchmark, such that in every round $t$, the investor chooses the best investment option that constitutes the optimal allocation and thus the optimal welfare for that round. Such a time-varying benchmark is \emph{stronger} than the benchmark in \cref{sec:approximation-dynamic-investment} where the investor $\iota$ sticks to a \textit{fixed} (or, more accurately, the \textit{best-in-hindsight}) investment under optimal allocation algorithm. We have shown that the approximation ratio $\beta$ of an XCONE allocation algorithm is preserved, with the regret defined on a fixed benchmark. However, it is impossible that the approximation ratio $\beta$ could still be preserved against the stronger benchmark. We provide further details below.

Recall that the \textbf{(expected) welfare} of an allocation algorithm $x$ at a dynamic investment instance $\tl{\omega}^T$ when the investor uses an online learning algorithm $\cL$ is defined as
        $$\tl{W}_x(\tl{\omega}^T,\cL)=\E_{s^{(t)}\sim g^{(t)},\cL}\left[\sum_{t=1}^T \ov{W}_x(\ov{\omega}^{(t)}, s^{(t)},\nu_\iota^{(t)},c^{(t)})\right].$$
\begin{definition}
    The \textbf{strongly dynamic (expected) welfare} at a dynamic investment instance $\tl{\omega}^T$ is
        $$\tl{W}^*_{\mathrm{dyn}}(\tl{\omega}^T)= \sum_{t=1}^T\max_{(\nu_\iota,c)\in I} \ov{W}^*(\ov{\omega}^{(t)},\nu_\iota,c).$$
\end{definition}
\begin{definition}
    An allocation algorithm $x$ is a \textbf{$\beta$-approximation for strongly dynamic investment}, if, for every dynamic investment instance $\tl{\omega}^T$ and an online learning algorithm $\cL$,
    \begin{align}
    \label{def:beta-dynamic-optimal}
    \tl{W}_x(\tl{\omega}^T,\cL) \ge \beta \tl{W}^*_{\mathrm{dyn}}(\tl{\omega}^T) -\Reg(\tl{\omega}^T,\cL).   
    \end{align}
\end{definition}

We show that even for allocation algorithms that are optimal in VCG mechanisms, a factor of $\frac{1}{|I|}$ on the approximation ratio is unavoidable for the stronger regret benchmark. \Cref{prop:1/I lower} characterizes such a lower bound for the stronger regret benchmark.
\begin{proposition}\label{prop:1/I lower}
    The approximation ratio of an optimal allocation algorithm 
    for strongly dynamic investment is at most $\frac{1}{|I|}$, where $|I|$ is the number of investments.
\end{proposition}
On the positive side, we show that if an allocation algorithm is $\beta$-approximation for dynamic investment, then it is also $\frac{\beta}{|I|}$-approximation for the strongly dynamic investment. Such an approximation ratio of $\frac{\beta}{|I|}$ matches the lower bound provided in \Cref{prop:1/I lower}.
\begin{proposition}\label{prop: 1/I upper}
    For any weakly monotone allocation algorithm $x$ and any $\beta \in [0,1]$, if $x$ is $\beta$-approximation for dynamic investment, then $x$ is also $\frac{\beta}{|I|}$-approximation for strongly dynamic investment.
\end{proposition}

Recall that any allocation algorithm with $\beta$-approximation for static investment is also $\beta$-approximation for static investment. That is no longer the case for the stronger time-varying benchmark. We show that it is unavoidable that any algorithm suffers a lower bound of $\frac{1}{|I|}$ on the approximation ratio. On the other hand, however, we provide a complete characterization of the approximation ratio for strongly dynamic investment by providing a matching upper bound $\frac{\beta}{|I|}$ for $\beta \in [0,1]$ for the lower bound of $\frac{1}{|I|}$. Therefore, the approximation ratio $\frac{1}{|I|}$ is both sufficient and necessary for strongly dynamic investment.



\section{Conclusion and Discussion}
\label{sec:conclusion}
\noindent

Achieving near-optimal social welfare assumes that the bidders always best respond in their participation in truthful mechanisms and that the investor, as one of the bidders among them, always selects the investment that is most favorable to her. This assumption is unrealistic in real-world resource allocation problems due to many reasons, e.g., the investor's limited information, the evolving environment and the investor's limited cognition for computing an optimal investment. 

In this paper, we study the preservation of a \textit{learning} investor's \textit{ex ante} investment incentives in a dynamic environment by adopting a more realistic behavioral assumption for the investor: no-regret learning. We ask the following question: \textit{What is the investment efficiency of a mechanism in a dynamic environment with an investor who learns to be no-regret?}

We consider a dynamic environment with repeated threshold auctions. For each round, the uncertainty of the environment is reflected by a state drawn from a certain distribution. The investor has no information about the exogenous environment, while she only knows her set of investments and the length of the learning periods. The investor may rely on an online learning algorithm for making investment decisions over time, which takes the investor's historical utilities as the input and helps with the investor's future investment decision-making.

Our main result is that the approximation ratio for the optimal welfare in a static environment, where the investor is required to best respond, still remains the same in a dynamic evolving environment, where the investor's best responses are not required. Additionally, we provide complete bounds for the approximation ratio for a stronger dynamic regret with the time-varying benchmark.

With close relationships between monotone algorithms and threshold auctions, as well as between Pareto efficiency and regret performance of online learning algorithms, our work is important, at the interface of mechanism design theory and learning in computer science, in providing insights into the efficiency of \textit{ex ante} investment incentives in mechanism design.

For open questions, future works may follow up on the consideration of multiple investors---what will happen when there is more than one investor who uses no-regret learning algorithms in their simultaneous participation of the repeated threshold auctions? Without learning, with multiple investors, \cite{AKLLM23} show that there will be inefficient equilibria even in the VCG mechanisms. With learning, will the learning help multiple investors avoid such inefficiency? Moreover, will the same approximation ratio in the dynamic environment with a single investor still be preserved when there are multiple learning investors? Meanwhile, we remark that some equilibria are inefficient even in VCG mechanisms with investors who are able to make best responses. Thus, one may expect to construct new models, possibly with new assumptions, to address the multiple investors' concerns, as our current model cannot be extended to the consideration of multiple investors. 


\newpage
\bibliographystyle{apalike}
\bibliography{main}

\newpage
\appendix
\section{Missing Proofs in \Cref{sec:stronger benchmark}}
\begin{proof}[Proof of \Cref{prop:1/I lower}]
    We consider a simple dynamic investment instance with only one investor and no other agents. We will construct an instance such that in each iteration, there is only one random investment option that results in a social welfare of $1$, while other options all result in a social welfare of $0$. Thus, the optimal welfare is 1, while any algorithm can only make a random guess between $|I|$ investments and get an expected welfare of $\frac{1}{|I|}$.
    
    Let $S = \{s_i\}_{i \in I}$ be the set of states. The investments are from the set $I = \{(v_i, 0)\}$ where $v_i : S \rightarrow \mathbb{R}$ is the indicator function such that $v_i(s_i) = 1$ and $v_i(s_j) =0$ for all $i \ne j \in I$.  The allocation outcome is binary, with the set of allocation outcomes being $\{0,1\}$. We have only one item for allocation. There is no constraint (e.g., the size constraint) for the allocation. Clearly, the optimal allocation outcome is that the investor always gets the allocation. Now consider the following distribution of dynamic investment instances $\tl{\omega}^T$ where the state distribution $g^{(t)}$ is uniformly at random drawn from $\{1_{s_i}\}_{s_i \in S}$ for all $t \in [T]$. Then we know the strongly dynamic social welfare is $$\tl{W}^*_{\mathrm{dyn}}(\tl{\omega}^T)= \sum_{t=1}^T\max_{(\nu_\iota,c)\in I} \ov{W}^*(\ov{\omega}^{(t)},\nu_\iota,c) = T.$$
    However, any algorithm $\cL$'s expected social welfare is only
    \begin{equation}
       \tl{W}_x(\tl{\omega}^T,\cL)=\E_{s^{(t)}\sim g^{(t)},\cL}\left[\sum_{t=1}^T \ov{W}_x(\ov{\omega}^{(t)}, s^{(t)},\nu_\iota^{(t)},c^{(t)})\right] = \frac{T}{|I|}.
    \end{equation}
\end{proof}

\begin{proof}[Proof of \Cref{prop: 1/I upper}]
    By the definition of $\beta$-approximation for dynamic investment, we have
    \begin{align}
        \tl{W}_x(\tl{\omega}^T,\cL) &\ge \beta \tl{W}^*(\tl{\omega}^T) - \Reg(\tl{\omega}^T,\cL) \\
        &= \beta \max_{(\nu_\iota,c)\in I} \sum_{t=1}^T \ov{W}^*(\ov{\omega}^{(t)},\nu_\iota,c) - \Reg(\tl{\omega}^T,\cL) \\
        & \ge \frac{\beta}{|I|} \sum_{(\nu_\iota,c)\in I}\sum_{t=1}^T \ov{W}^*(\ov{\omega}^{(t)},\nu_\iota,c) - \Reg(\tl{\omega}^T,\cL) \\
        & \ge \frac{\beta}{|I|} \sum_{t=1}^T\max_{(\nu_\iota,c)\in I} \ov{W}^* (\ov{\omega}^{(t)},\nu_\iota,c) - \Reg(\tl{\omega}^T,\cL) \\
        & =\frac{\beta}{|I|} \tl{W}^*_{\mathrm{dyn}}(\tl{\omega}^T)- \Reg(\tl{\omega}^T,\cL). 
    \end{align}
    This completes the proof.
\end{proof}

\end{document}